\documentclass[prb,twocolumn,showpacs,aps,amssymb,superscriptaddress]{revtex4-1}
\usepackage{graphicx}
\usepackage{epsfig}
\usepackage{color}
\usepackage{ulem}
\newcommand{\nn}[1]{{\langle #1 \rangle}}

\begin{document}

\title{Supersolid polar molecules beyond pairwise interactions}
\author{Lars Bonnes}
\affiliation{Institut f\"ur Theoretische Physik III,Universit\"at Stuttgart, Pfaffenwaldring 57, 70550 Stuttgart, Germany}
\author{Stefan Wessel}
\affiliation{Institut f\"ur Theoretische Physik III,Universit\"at Stuttgart, Pfaffenwaldring 57, 70550 Stuttgart, Germany}
\affiliation{Institute for Theoretical Solid State Physics, RWTH Aachen University, Otto-Blumenthal-Str. 26, 52056 Aachen}
\begin{abstract}
We explore the phase diagram of ultracold bosonic polar molecules confined to a planar 
optical lattice of triangular geometry.
External static electric and microwave fields can be employed to tune  the effective interactions
between the polar molecules into a regime of extended two- and three-body repulsions of comparable strength, leading to a rich quantum phase diagram.
In addition to various solid phases, an extended supersolid phase is found to persist deep into the three-body dominated regime. 
While three-body interactions  break particle-hole symmetry explicitly, a characteristic supersolid-supersolid quantum phase transition is observed, 
which indicates the restoration of particle-hole symmetry at half-filling. We revisit the spatial structure of the supersolid at this filling,  regarding  the existence of a
further supersolid phase with three inequivalent sublattices, and provide evidence that this state is excluded also at finite temperatures.

\end{abstract}
\maketitle
\section{Introduction}
Identifying  conditions for  bulk supersolidity, both in the continuum and for bosons confined in periodic potentials, 
is a major task in order to explore the supersolid state of matter.~\cite{penrose56, andreev69, legget70}
A fascinating approach towards probing the range of stability and the nature of supersolids would be provided by the direct manipulation of the inter-particle interactions. 
In this respect, ultracold polar molecules~\cite{sage05, ni08, deiglmayr08, lahaye09} exhibit the remarkable opportunity to tailor  many-body interactions by the application of static electric and microwave fields.~\cite{micheli06, wang06, kotochigova06}
It has been demonstrated, that by appropriately tuning  the external control parameters, polar molecules can in fact  be driven into a regime, where  three-body repulsions strongly compete with effective two-body interactions between the particles.~\cite{buechler07} Starting from a supersolid phase,  this tune-ability thus allows to explore the stability of lattice supersolid states beyond the regime of pairwise (two-body)  interactions. 

Here, we consider in particular a two-dimensional setup, where the molecular dipole moments 
align parallel to a  static electric field directed perpendicular to the confining plane. The effective dynamics of the polar molecules in an optical lattice potential is then described 
by a Bose-Hubbard Hamiltonian of hard-core bosons.~\cite{buechler07}
\begin{equation}
H =	- t \sum_\nn{ij} b_i^\dagger b_j 
 + \sum_{ij} \frac{V_{ij}}{2} n_i n_j+ \sum_{ijk} \frac{W_{ijk}}{6} n_i n_j n_k-\mu \sum_{i} n_i,
\end{equation}
where 
$b_i$ and $b_i^\dagger$ denote boson annihilation and creation operators respectively on lattice site $i$, and the local density operator $n_i=b_i^\dagger b_i$. Furthermore, $t$ denotes the nearest-neighbor hopping matrix element, and $\mu$ the chemical potential, controlling the filling $n$ of the lattice. In contrast to two-body terms, the three-body interactions explicitly break the particle-hole symmetry that the model exhibits in the limit $W=0$.
The interactions furthermore depend on the lattice geometry via their respective strengths, 
\begin{equation}
V_{ij}=V/r^6_{ij}
\end{equation}
and
\begin{equation}
W_{ijk}=W\left[\frac{1}{r^{3}_{ij}r^{3}_{ik}}+\frac{1}{r^{3}_{ij}r^{3}_{jk}}+\frac{1}{r^{3}_{ik}r^{3}_{jk}}\right],
\end{equation}
where $r_{ij}$ denotes the distance between lattice sites $i$ and $j$. 
We consider here the case of a triangular lattice, for which 
the geometric factors  of the leading contributions to the interaction terms are given in Fig.~1.  
Analyzing the dependence of these parameters on the experimental setup shows
that $W/V$ can be tuned up to about $0.7$.~\cite{buechler07} Furthermore, $t$ can be varied independently by controlling the depth of the optical lattice.

\begin{figure}[t]
\begin{center}
\includegraphics[width=8.5cm]{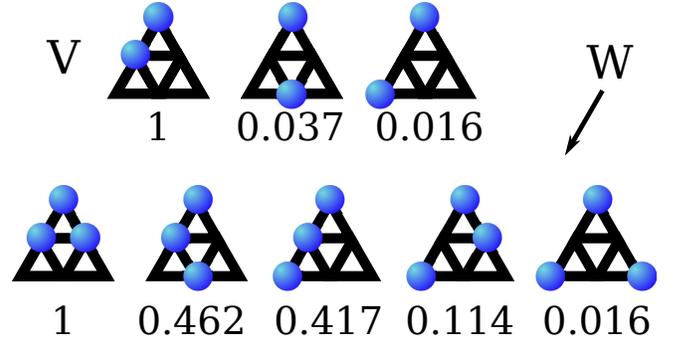}
\caption{(Color online) Leading contributions to the many-body interactions for 
polar molecules on a triangular optical lattice. 
Numbers below the interaction terms denote the relative strength of the two- (three-) body terms in units of the nearest-neighbor values $V$ ($W$).
}
\end{center}
\end{figure}

We explore the consequences of three-body interactions on the low-energy properties. Of particular interest are the stability and  nature of the supersolid (SS) state  of  bosons on the triangular lattice found in the particular cases of nearest-neighbor~\cite{murthy97,wessel05a,heidarian05,melko05,boninsegni06} as well as for   $1/r_{ij}^3$-extended two-body repulsions.~\cite{pollet10}
For the nearest-neighbor case, a discontinuous quantum phase transition separates two distinct SS phases, which differ in their internal structure.~\cite{boninsegni06} In the SS-A (SS-B) phase, of density $n$ above (below) 1/2, two sublattices of the triangular lattice show density fluctuations about a  larger (smaller) common mean value than the sites of the third sublattice. These phases are related by the exchange of particles and holes, and
meet at the particle-hole symmetric line of half-filling, where another possible SS state (SS-C), with distinct occupations on all three sublattices,~\cite{melko05,hassan07} is indeed excluded in the ground state by the global $U(1)$ symmetry of particle number conservation.~\cite{boninsegni06}

\begin{figure}[t]
\begin{center}
\includegraphics[width=7.5cm]{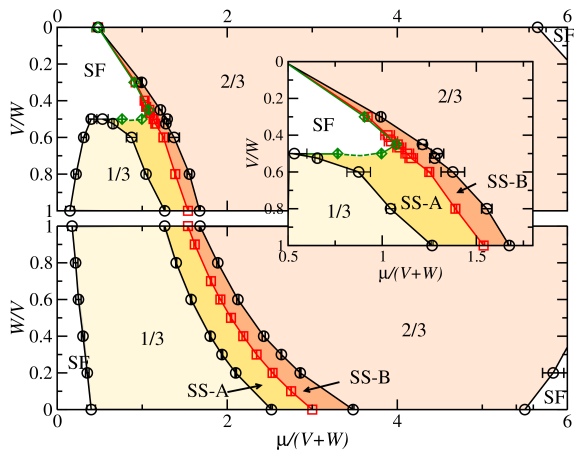}
\caption{(Color online) Ground state phase diagram of 
polar molecules on a triangular lattices. 
The upper (lower) panel shows the regime $W>V$ ($V>W$), at $t/W=0.1$ ($t/V=0.1$).
SF denotes superfluid regions, SS-A and SS-B the two supersolid regimes, that 
meet at a first-order quantum phase transition along the line  of half-filling, $n=1/2$ (squares). Diamonds trace the SS-SF transition line. 
The inset focuses in on the large-$W$ region.
}
\end{center}
\end{figure}

In order to study the effects of three-body repulsions on the SS phases,  
we first restrict to the nearest-neighbor two- and three-body terms  in $H$, i.e. the most basic model with competing two- and three-body interactions on the triangular lattice. Then, we assess the effects of longer-ranged interactions. 
The stability of the  SS phase under three-body interactions is observed already within the reduced Hamiltonian. This is unlike the case of the square lattice,~\cite{schmidt08}
for which the emergence of a SS state relies on interaction ranges beyond nearest-neighbors.


\section{Method}
We employed quantum Monte Carlo simulations using a generalized directed-loop algorithm in the stochastic series expansion representation.~\cite{sandvik99b,syljuasen02,alet05} Decoupling the Hamiltonian in terms of 6-sites clusters as in Fig.~1, allows us to account efficiently for all interactions of strengths larger than $0.016V$, and $0.05W$, respectively. The simulations were performed on finite systems with periodic boundary condition, containing $N=L^2$ lattice sites ($L$ denotes the linear system size), with $L$ up to 48. In order to obtain ground state properties for these finite systems, the temperature $T$ needs to be chosen sufficiently low, as specified below. 

\section{Phase diagram}
The ground state phase diagram for the nearest-neighbor model is shown in Fig. 2 in the vicinity of the supersolid phase near half-filling and for $t=0.1 \max(W,V)$. Besides the superfluid (SF) regime two solid phases appear of densities $n=1/3$ and $2/3$, respectively. The corresponding density structure factor 
\begin{equation}
S=\frac{1}{N} \left<  \sum_{i,j} e^{i{\mathbf Q}({\mathbf r_i}-{\mathbf r_j})} n_i n_j \right>
\end{equation}
relates to the density correlation function at the corner of the hexagonal Brillouin zone at momentum ${\mathbf Q} \equiv (2\pi/3,0)$.~\cite{wessel05a}
In addition, the phase diagram exhibits an extended SS region, where both $S/N$ and the SF density $\rho_S$ scale to finite values in the thermodynamic limit. Here,  
\begin{equation}
\rho_S=\frac{T}{2t}\langle W^2\rangle
\end{equation}
is obtained from measuring the boson winding number $W$ fluctuations.~\cite{pollock87} The phase boundaries in Fig. 2 result from data such as those shown in Fig. 3, after an extrapolation to the thermodynamic limit (cf. the inset of Fig. 3 for an example). 

\begin{figure}[t]
\begin{center}
\includegraphics[width=8cm]{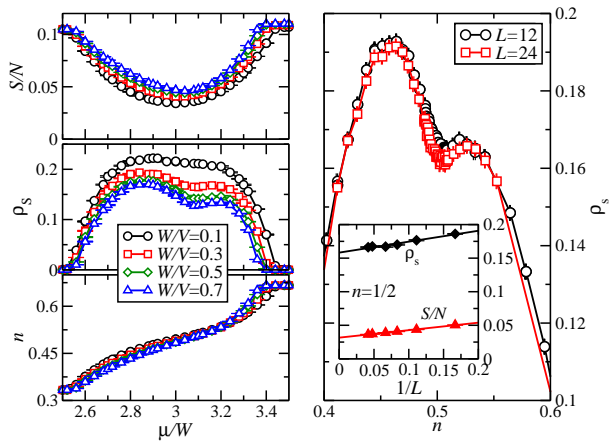}
\caption{(Color online) {\it Left:}
Filling $n$, superfluid density $\rho_S$, and density
structure factor $S$ as a function of the chemical potential $\mu$ for
various values of $W/V$ at $t/V=0.1$, taken for $L=12$ and $T/t=0.1$.
{\it Right:} Filling $n$ dependence of the superfluid density $\rho_S$ at $W/V=0.3$ and $t/V=0.1$, taken for $L=12$ 
(squares) and $24$ (solid circles) and $T/t=0.1$. 
Inset: Finite-size extrapolation of $\rho_S$ (circles) and $S/N$ (squares) at $n=1/2$.
}
\end{center}
\end{figure}
We find, that the SS state, realized on the triangular lattice in the absence of three-body interactions, remains stable even for large values of $W/V$. Only in the limit of purely three-body interactions ($V=0$) is the SS destroyed. The solid phase of filling $n=1/3$ is stable up to $W\approx 2V$. This destabilization of the $n=1/3$ solid phase via $W$ is 
consistent with results obtained for the classical limit ($t=0$) of our model, i.e. the Ising model with two- and three-body interactions in a magnetic field.~\cite{schick77,chin97}

\begin{figure}[t]
\begin{center}
\includegraphics[width=8cm]{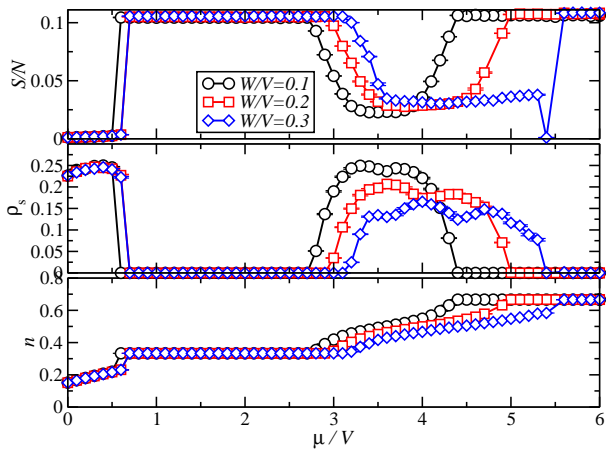}
\caption{ (Color online)
Filling $n$, Superfluid density $\rho_S$, and the density
structure factor $S$ as functions of the chemical potential $\mu$ for
different values of $W/V$ at $t/V=0.1$, taken for $L=12$ and $T/t=0.1$ in the extended model.}
\end{center}
\end{figure}

\section{Extended model}
We now turn to the extended model approximating the long-range nature of the dressed dipolar interactions by taking into account all interaction vertices drawn in Fig.~1.
First, we consider the parameter region, where the SS phase is stabilized. In particular, 
Fig.~4 shows $n$, $\rho_S$ and $S$ as functions of $\mu$ at $t/V=0.1$ for different ratios $W/V<1$ around the SS regime. 
An extended SS phase is clearly identified between the $n=1/3$ and the $n=2/3$ solid. 

Next, we explore the phase diagram of the extended model beyond the SS region. Of particular interest are  incompressible phases that  are stabilized by the three-body interactions. To explore these,  we first consider the limiting case of purely three-body interactions  ($V=0$).
Fig.~5 shows quantum Monte Carlo data of the filling  as a function of $\mu$ for different ratios  $t/W$ at $V=0$.
In addition to the previously identified density plateaus at fillings $n=1/3$ and $n=2/3$, 
incompressible  phases at fillings $n=3/8$, $1/2$, and $3/4$ are  seen to emerge upon decreasing $t/W$. The spatial structures that characterize these solids are shown in Fig.~6.

In the $n=3/8$ solid, the system forms irregular patterns based on nearest neighbor dimers. From inspecting various real space density distributions from the quantum Monte Carlo configurations, 
we conclude that these dimers are not arrange in any periodic pattern. 
This dimerization leads to a strong energy penalty for finite values of $V$, such that this phase is unstable for $V>W$. 
Similarly, the $n=1/2$ phase, where   stripes of empty and occupied lattice sites alternate, is not observed within this  parameter regime. 
The solid phase at $n=3/4$ however remains stable; for example we observe it for $W/V=0.2$, $t/V=0.1$ around $\mu/V=14$ (not shown).

Due to the frustration induced be the competition between the two- and three-body interaction terms, 
we are able to perform ergodic simulations for the extended model only for ratios $W/V$ up to about $0.3$. For example, as the dip in the data for $S/N$ near $\mu/V=5.5$ in Fig.~4 illustrates, 
we cannot reliable resolve the transition region between the SS phase and the $n=2/3$ solid for the extended model. 
Using annealing techniques, the simulations did not improve within the transition region. 
It remains possible, that another, maybe incommensurate, phase appears here. However, from our simulations, we are not able to conclude on this issue. 
Multi-body interactions at larger distances, as seen above for the case of purely three-body interactions, are in fact expected to stabilize additional plateaus at further fractional fillings, 
with the possible  proliferation of low-energy metastable states~\cite{menotti07} and Spivak-Kivelson bubble-type transitions between solid and superfluid regions.~\cite{spivak04} 
However, a reliable numerical study of these issues is  beyond the present approach.~\cite{pollet10}

\begin{figure}[t]
\begin{center}
\includegraphics[width=8cm]{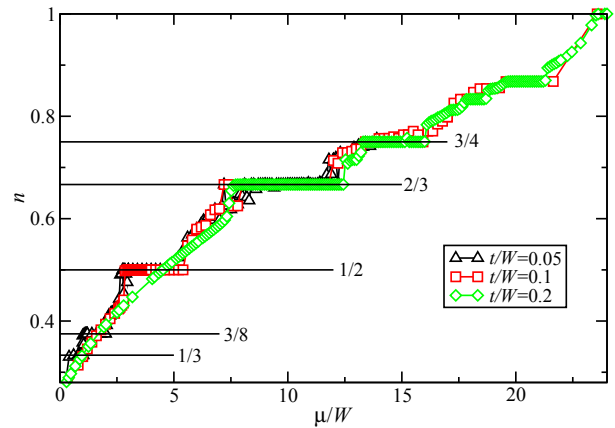}
\caption{(Color online)
Density $n$ as a function of the chemical potential $\mu$ within the extended model
for different values of $t/W$ at $V=0$, taken for $L=12$ and $T/t=0.1$.
}
\end{center}
\end{figure}

\begin{figure}[t]
\begin{center}
\includegraphics[width=8cm]{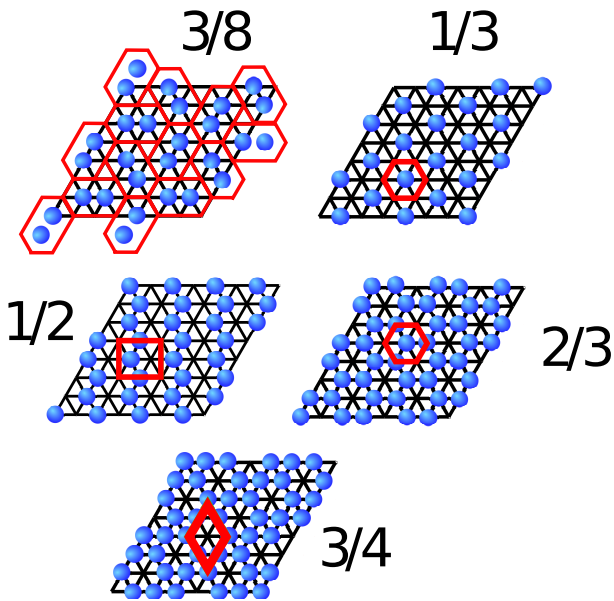}
\caption{(Color online)
Real-space structures of incompressible phases stabilized in the extended model, labeled by their filling fraction $n$. For crystalline states, the crystal unit cell is also indicated.
For the $n=3/8$ phase, which does not exhibit a periodic structure, a typical state is shown,  consisting of an irregular arrangement of isolated dimers (pairs of neighboring occupied sites).
}
\end{center}
\end{figure}

\section{SS-SS transition}
\begin{figure}[t]
\begin{center}
\includegraphics[width=9cm]{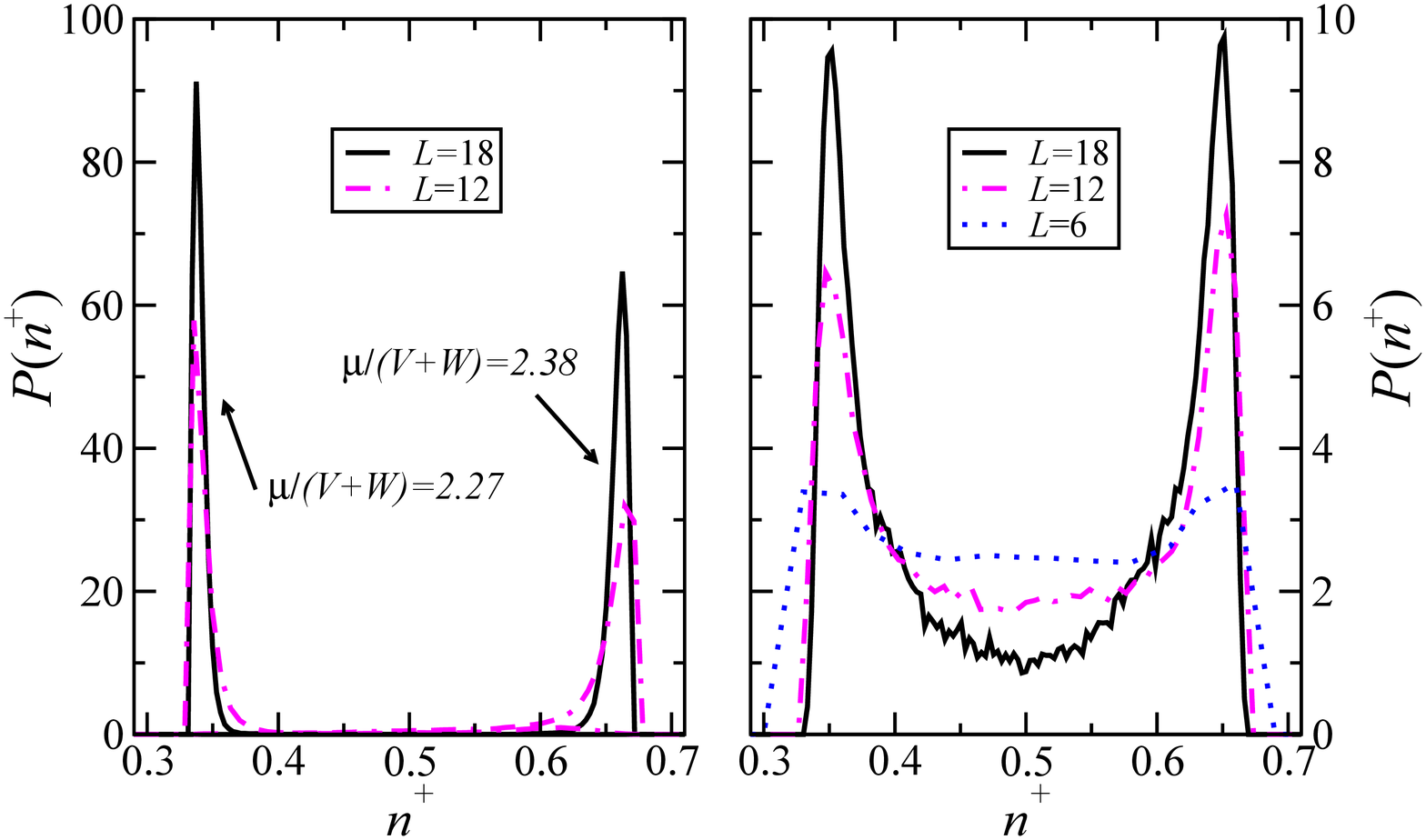}
\caption{(Color online) 
Histogram $P$ of $n^+$ for different system sizes taken at $t/V=0.1$, $W/V=0.3$, and $T/t=0.05$ within the SS-A and SS-B phase (left panel) and at the SS-SS transition at $n=1/2$ (right panel).
}
\end{center}
\end{figure}

\begin{figure}[t]
\begin{center}
\includegraphics[width=8cm]{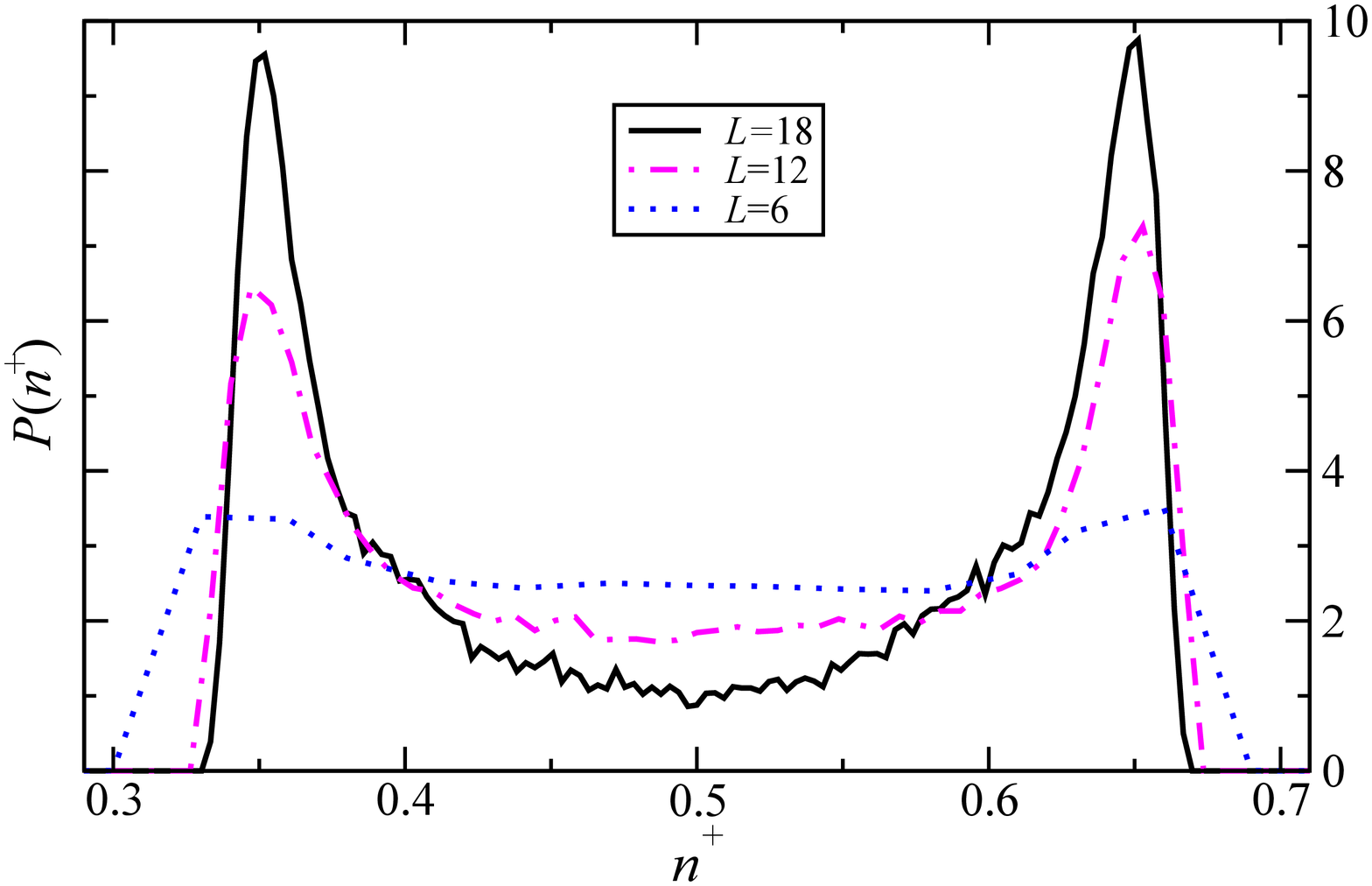}
\caption{ (Color online)
Histograms $P$ of $n^+$ for different system sizes taken for the extended model at $W/V=0.1$, $t/V=0.1$ and $T/t=0.1$ at half-filling $n=1/2$.
}
\end{center}
\end{figure}
In the curves of the superfluid density as functions of $\mu$ in Fig.~3 (left panel), dips occur for values of $\mu/V$ near $3.1$. 
This suppression of $\rho_S$ appears around $n=1/2$, as seen from Fig.~3 (right panel), which shows $\rho_S$ as a function of the filling $n$.  
As seen from the finite size extrapolation in the inset of Fig.~3, the SS however remains stable in the thermodynamic limit at this filling.
The suppression of $\rho_S$ relates to 
a remarkable feature of supersolidity of hard-core bosons on the triangular lattice: the presence of a SS-A to SS-B quantum phase transition at half-filling.~\cite{boninsegni06}
A robust means of identifying this SS-SS transition is to monitor the histogram of the local densities.~\cite{boninsegni06} 
Denoting by  
\begin{equation}
\bar{n}_i=\beta^{-1}\int_0^\beta d\tau n_i(\tau)
\end{equation}
the imaginary-time averaged local density at lattice site $i$ for a given Monte Carlo configuration, 
one obtains the fraction of sites with mean occupation $\bar{n}_i>1/2$ from the observable
\begin{equation}
n^+=\frac{1}{N} \sum_{i=1}^N \theta(\bar{n}_i-1/2),
\end{equation}
where $\theta$ denotes the Heaviside function. 
Supersolid ground states belonging to the  SS-A (SS-B) type are characterized by  peaks in a histogram $P$ of $n^+$ near 2/3 (1/3), such as those in the left panel of Fig.~7. 
The observation of a doubly-peaked histogram in the quantum Monte Carlo data exhibits the presence of both SS-A as well as SS-B configurations at $n=1/2$, whereas the SS-C phase would correspond to detecting a singly-peaked histogram at $n^+=1/2$. 
Performing such an identification of the nature of the SS state for $W\neq 0$, we find that the suppression in $\rho_S$ is due  to the SS-SS quantum phase transition at $n=1/2$.
As an example, we show in the right panel of  Fig.~8 histograms taken for different system sizes at $n=1/2$ for $W/V=0.3$ and $t/V=0.1$. One clearly identifies for sufficiently large system sizes a characteristic two-peak structure, indicative of the SS-A, SS-B coexistence at half-filling. 
This feature is not restricted to the nearest-neighbor model. In Fig.~4, for the extended model, similar dips are observed in $\rho_S$ as for the nearest-neighbor case.
The histogram of $n^+$, measured for the extended model at $W/V=0.1$ and $t/V=0.05$ and shown in Fig.~8, indeed also exhibits the characteristic double-peak structure of the SS-A, SS-B coexistence. 
Hence, although particle-hole symmetry is explicitly broken for $W\neq 0$, the characteristic SS-SS transition indicates its dynamical restoration in the SS region along the line of half-filling, $n=1/2$. 

While the $n=1/3$ solid is not stable for large values of $W/V\gtrsim 2$ (cf. the inset of Fig.~2), we observe that within the remaining SS stripe both the SS-A and the SS-B phases remain stable. 
In fact,  upon increasing $\mu$ at $V/W=0.3$, one enters the SS region from the SF region at  $n=0.48(1) < 1/2$, i.e. into the SS-A phase, before the transition to the SS-B phase takes place at $n=1/2$. 
This stability of the SS-A phase furthermore leads to a SS re-entrance phenomenon when  the density increases at a fixed ratio of $V/W=0.5$:
{Fig.~9 shows the superfluid density $\rho_s$ and the structure factor $S/N$ at $V/W=0.5$ and $t/W=0.1$. 
Upon increasing the chemical potential $\mu/W$, 
the system's ground state changes (i)
from the 1/3-solid to a SS-A state at $\mu/W\approx 0.9$, then (ii) 
becomes a uniform superfluid at $\mu/W \approx 1.2$, before (iii) a transition occurs around
 $\mu/W \approx 1.5$ to the SS-A phase, which eventually (iv) at $\mu/W \approx 1.9$ enters into the 2/3-solid phase. 
In particular, we thus observe here re-entrance behavior with respect to the SS-A phase upon increasing the filling.
This is indicated by the dashed line in Fig.~2.

\begin{figure}[t]
\begin{center}
\includegraphics[width=8cm]{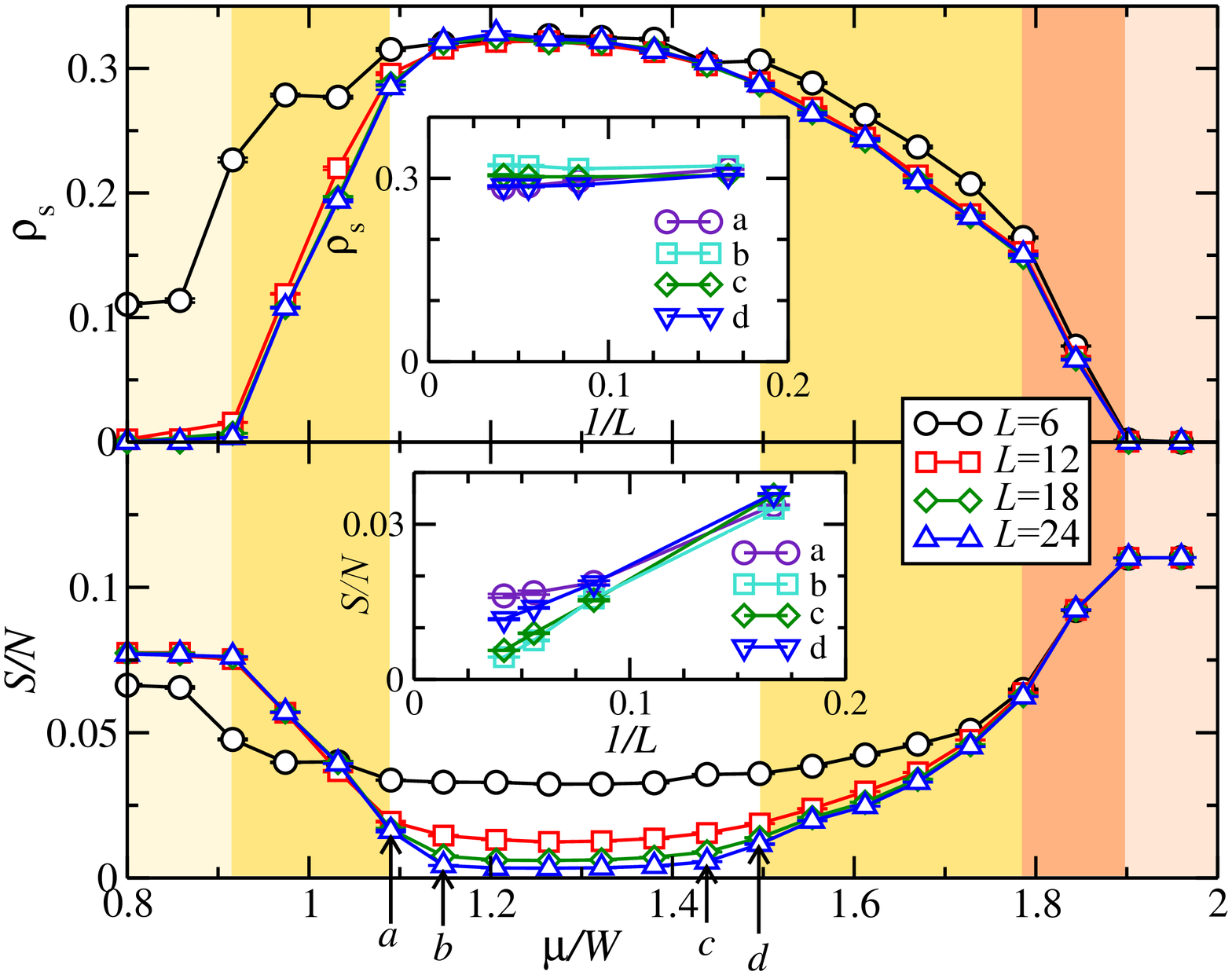}
\caption{(Color online)
Superfluid density $\rho_S$ (upper panel) and structure factor $S$
(lower panel) at $V/W=0.5$, $t/W=0.1$ for different system sizes.
The different phases are highlighted (light) yellow for (the low density solid) SS-A and (light) red for (the high density solid) SS-B, respectively.
The inset shows finite size extrapolations for the structure factor at $\mu/W=1.09$ (a), 1.148(b), 1.438 (c), and 1.496 (d).}
\end{center}
\end{figure}

\section{Assessing the SS-C Phase}
{A recent self-consistent cluster mean-field approach reports that the
SS-C phase can be stabilized by thermal fluctuation.~\cite{hassan07} At $n=1/2$, the SS-C state was obtained at $t/V=0.1$ for $T/t>0.5$, while the SS melting temperature was estimated near $T/t=6$. This temperature regime is not consistent with the results in Ref.~16, where it was found, that the SS melts already around $T/t=0.5$. We thus performed finite-temperature simulations in order to assess the proposed SS-C scenario.
In our histograms (c.f. Fig.~10) at $W=0$ and $t/V=0.1$ a double-peak structure and a suppression at $n^+=1/2$ emerged for sufficiently large system sizes at least up to $T/t\approx 0.25$, indicative of the SS-A and SS-B coexistence at $n=1/2$. The peaks broaden and shift towards $n^+=1/2$ upon increasing $T$. For $T/t$ beyond 0.25, we cannot resolve a double-peak structure, instead a single broad peak  around $n^+=1/2$ persists up to the largest system size considered.
Since we find that  thermal disorder  leads to a substantial weight near $n^+=1/2$,  this observation does however not provide evidence for a SS-C state, but instead suggests that the double-peak structure is beyond our numerical resolution for $T/t$ beyond  $0.25$. Within quantum Monte Carlo, the SS-C thus remains elusive. We also did not observe it for $W\neq 0$.}

\begin{figure}[t]
\begin{center}
\includegraphics[width=8cm]{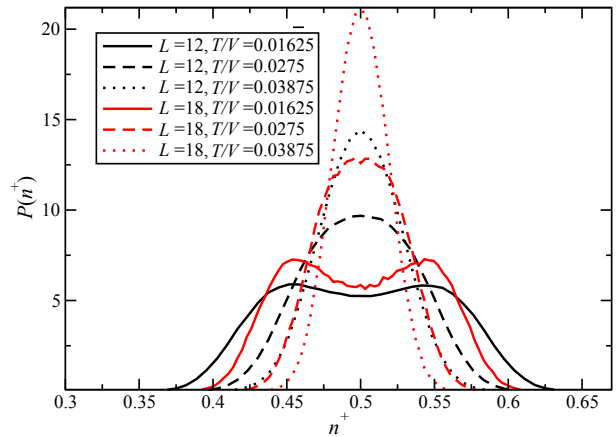}
\caption{
Histograms $P$ of $n^+$  taken at $W=0$, $t/V=0.1$, $\mu/V=3$ for different temperatures and system sizes.
}
\end{center}
\end{figure}

\section{Conclusions}
We showed that polarized polar molecules on the triangular lattice exhibit a robust supersolid phase in the presence of strong three-body interactions. 
The broken particle-hole symmetry is recovered along the line of half-filling, as indicated by the persistence of a first-order supersolid-supersolid transition. 
Furthermore, based on our findings, we exclude the existence of a SS-C phase~\cite{boninsegni06} also at finite temperatures.
Experimentally,
the difference between a normal liquid, commensurate solids and supersolid phases can be detected via time-of-flight images. 
Density correlations can also be probed using Bragg spectroscopy.~\cite{papp08}
Furthermore, 
detecting the spatial structure of the solid and supersolid phases is achieved with a local addressability of individual lattice sites; the feasibility of this method for atoms has indeed been demonstrated recently.~\cite{bakr09,bakr10,sherson10}
For the future, we plan to analyze in detail the quantum phase transition between the supersolid and the superfluid phase, for which a recent cluster 
mean-field theory~\cite{yamamoto11} predicts that
for the nearest-neighbor model this transition is first-order  away from the particle-hole symmetric point at $\mu/V=3$, where instead a continuous transition was obtained, consistent with previous findings.~\cite{wessel05a,melko05}
It will be interesting to 
verify this prediction and to clarify the unversality class of the later continuous transition.

\begin{acknowledgements}
We thank H.-P. B\"uchler, A.M. L\"auchli and K.P. Schmidt 
for discussions, and
acknowledge the allocation of CPU time on the HLRS Stuttgart 
and NIC J\"ulich supercomputers. 
Support was also provided through 
the Studienstiftung des Deutschen Volkes (LB)
and
the DFG within SFB/TRR 21 (SW).
\end{acknowledgements}


\begin{thebibliography}{10}
\bibitem{penrose56}
O. Penrose and L. Onsager, Phys. Rev. {\bf 104},  576  (1956).
\bibitem{andreev69}
A.~F. Andreev and I.~M. Lifshitz, Sov. Phys. JETP {\bf 29},  1107  (1969).
\bibitem{legget70}
A.~J. Leggett, Phys. Rev. Lett. {\bf 25},  1543  (1970).

\bibitem{sage05}
J.~M. Sage, S. Sainis, T. Bergeman, and D. DeMille, Phys. Rev. Lett. {\bf 94},
  203001  (2005).
\bibitem{ni08}  
K.-K. Ni, S. Ospelkaus, M. H. G. {de Miranda}, A. Pe'er, B. Neyenhuis, J. J. Zirbel, S. Kotochigova, P. S. Julienne, D. S. Jin, and J. Ye, Science {\bf 322},  231  (2008).
\bibitem{deiglmayr08}
J. Deiglmayr, A. Grochola, M. Repp, K. Mortlbauer, C. Gluck, J. Lange, O. Dulieu, R. Wester, and M. Weidem{\"u}ller, Phys. Rev. Lett. {\bf 101},  133004  (2008).
\bibitem{lahaye09}
T. Lahaye, C. Menotti, L. Santos, M. Lewenstein, and T Pfau, Rep. Prog. Phys. {\bf 72},  126401 (2009).

\bibitem{micheli06}
A. Micheli, G.~K. Brennen, and P. Zoller, Nature Physics {\bf 2},  341  (2006).
\bibitem{wang06}
D.-W. Wang, M.~D. Lukin, and E. Demler, Phys. Rev. Lett. {\bf 97},  180413 (2006).
\bibitem{kotochigova06} 
S. Kotochigova and E. Tiesinga, Phys. Rev. A {\bf 73},  041405  (2006).

\bibitem{buechler07}
H.~P. B{\"u}chler, A. Micheli, and P. Zoller, Nature Physics {\bf 3},  726 (2007).

\bibitem{murthy97}
G. Murthy, D. Arovas, and A. Auerbach, Phys. Rev. B {\bf 55},  3104  (1997).

\bibitem{wessel05a}
S. Wessel and M. Troyer, Phys. Rev. Lett. {\bf 95},  127205  (2005).

\bibitem{heidarian05}
D. Heidarian and K. Damle, Phys. Rev. Lett. {\bf 95},  127206  (2005).

\bibitem{melko05}
R. G. Melko, A. Paramekanti, A. A. Burkov, A. Vishwanath, D. N. Sheng, and L. Balents, Phys. Rev. Lett. {\bf 95},  127207  (2005).

\bibitem{boninsegni06}
M. Boninsegni, N. Prokof'ev, and B. Svistunov, Phys. Rev. Lett. {\bf 96}, 105301  (2006).

\bibitem{pollet10}
L. Pollet, J.~D. Picon, H.P. B{\"u}chler, and M. Troyer, Phys. Rev. Lett. {\bf
  104},  125302  (2010).

\bibitem{hassan07}
S.~R. Hassan, L. {de Medici}, and A.-M.~S. Tremblay, Phys. Rev. B {\bf 76},
  144420  (2007).

\bibitem{schmidt08}
{K. P. Schmidt, J. Dorier, and A. M. L{\"a}uchli}, Phys. Rev. Lett. {\bf 101},
  150405  (2008).

\bibitem{sandvik99b}
A.~W. Sandvik, Phys. Rev. B {\bf 59},  R14157  (1999).
\bibitem{syljuasen02}
O.~F. Sylju{\aa}sen and A.~W. Sandvik, Phys. Rev. E {\bf 66},  046701  (2002).
\bibitem{alet05}
F. Alet, S. Wessel, and M. Troyer, Phys. Rev. E {\bf 71},  036706  (2005).

\bibitem{pollock87}
E.~L. Pollock and D.~M. Ceperley, Phys. Rev. B. {\bf 36},  8343  (1987).

\bibitem{schick77}
M. Schick, J. S. Walker, and M. Wortis, Phys. Rev. B {\bf 16}, 2205 (1977).
\bibitem{chin97}
K. K. Chin and D. P. Landau, Phys. Rev. B {\bf 36}, 275 (1987).

\bibitem{menotti07}
C. Menotti, C. Trefzger, and M. Lewenstein, Phys. Rev. Lett. {\bf 98},  235301
  (2007).

\bibitem{spivak04}
B. Spivak and S.~A. Kivelson, Phys. Rev. B {\bf 70},  155114  (2004).

\bibitem{papp08}
S.~B. Papp, J. M. Pino, R. J. Wild, S. Ronen, C. E. Wieman, D. S. Jin, and E. A. Cornell, Phys. Rev. Lett. {\bf 101},  135301  (2008).

\bibitem{bakr09}
W.~S. Bakr, J. I. Gillen, A. Peng, S. F{\"o}lling, and M. Greiner, Nature {\bf 462},  74  (2009).
\bibitem{bakr10}
W.~S. Bakr, A. Peng, M. E. Tai, J. Simon, J. I. Gillen,S. F{\"o}lling, L. Pollet, and M. Greiner, Science {\bf 329},  547  (2010).
\bibitem{sherson10}
J. F. Sherson, C. Weitenberg, M. Endres, M. Cheneau, I. Bloch, and S. Kuhr, Nature {\bf 467}, 68 (2010).

\bibitem{yamamoto11}
D. Yamamoto, I. Danshita, and C. A. R. S{\'{a}} de Melo, arXiv:1102.1317v1.

\end{thebibliography}

\end{document}